# Towards Blockchain-enabled Wireless Mesh Networks

Mennan Selimi
University of Cambridge and
Ammbr Research Labs
Cambridge, UK

Aniruddh Rao Kabbinale
University of Cambridge
Cambridge, UK

Anwaar Ali
University of Cambridge
Cambridge, UK

Leandro Navarro
UPC BarcelonaTech and
Ammbr Research Labs
Barcelona, Spain

Arjuna Sathiaseelan
University of Cambridge and
Ammbr Research Labs
Cambridge, UK

## ABSTRACT

Recently, mesh networking and blockchain are two of the hottest technologies in the telecommunications industry. Combining both can reformulate internet access and make connecting to the Internet not only easy, but affordable too. Hyperledger Fabric (HLF) is a blockchain framework implementation and one of the Hyperledger projects hosted by The Linux Foundation. We evaluate HLF in a real production mesh network and in the laboratory, quantify its performance, bottlenecks and limitations of the current implementation. We identify the opportunities for improvement to serve the needs of wireless mesh access networks. To the best of our knowledge, this is the first HLF deployment made in a production wireless mesh network.

## 1. INTRODUCTION

Network infrastructures are critical to provide local and global connectivity that enable access to information, social inclusion and participation for everyone. Local connectivity largerly relies on access networks. Wireless mesh networks (WMNs) are the access networks comprising of wireless nodes namely wireless mesh routers and wireless mesh clients. A client (irrespective of whether it is a mesh or a generic client) can access the Internet through a WMN [1].

Community networks are network infrastructure commons, built by citizens and organizations which pool their resources and coordinate their efforts, characterized by being open, free and neutral [5].

Community Mesh Networks (CMNs) are a special case of WMNs which are usually setup as a community network. The CMNs have been identified as one of the models contributing to connecting the next billion people that are still without the Internet access [9]. Guifi.net[1] is an example of such a community effort which is one of the biggest community networks in the world, with more than 34.000 participating routers.

The idea of CMNs, nobel as it seems, does not come with out its fair caveats. Since the nature of CMNs is peer-to-peer there are concerns related to trust among various participating peers and how to make this volunteer effort economically viable and sustainable as well [3].

As an example scenario we consider the economic *compensation system* in Guifi.net [3]. The idea of the compensation system is to create a balance between total resource contribution and its consumption. The economic value of the contribution and consumption of network resources for each participant and in a given locality are recorded. The overall result is a zero-sum budget where the participants with over-consumption or negative balances, have to compensate those with over-contribution or positive balances.

Currently the above described economic compensation system is implemented in a manual manner. What this means is that a participant puts forward a claim of its consumption and then the Guifi.net foundation[2] validates this claim by cross checking it with their own network traffic measurement data and network inventory. Any disparities between these two records are then flagged. There is, however, room for errors or even malicious activities such as false claims put forward by a participant, the recorded data being tampered with, or simply mistrust among the parties. There is a need for a system where participants can trust that the consumption of resources is being accounted in a fair manner, and that these calculations and money transfers are done automatically to avoid the cost, delays, errors and potential mistrust from manual accounting and external payments.

Blockchain is one of the solutions that seems quite apt to make the peer-to-peer nature of CMNs trusted and economically sustainable. Blockchain (more details in Section 2) is an immutable and distributed data storage without the provision of retrospective mutation in data records. However, most blockchain networks are usually open and permissionless that also encourage the users of such a network to be anony-mous [10]. This implies that anyone, without revealing its true identity, can be part of such a network and make transactions with another similar anonymous peer of the network.

[1]`https://guifi.net/`

[2]`https://fundacio.guifi.net/Foundation`

In the perspective of CMNs, however, such as in Guifi.net every participant who joins the network to contribute to the infrastructure must first register its identity and the identity of the resources that it contributes to the wider pool. This is particularly needed so that a malicious entity such as hidden nodes in Guifi.net used by other ISPs to provide services to their users can be filtered out [11]. Because of such registration process one also needs an efficient identity mechanism on top of blockchain's immutable record keeping. *Permissioned blockhains* are part of such solutions. These blockchain solutions are mostly envisioned for business networks where there is often a stringent requirement of *know your customer* in addition to keeping the intra- and inter-business transactions confidential. Hyperledger[3] (see Section 2.1 for details) is one such solution that realises the concept of permissioned blockchains and which we also use in our current study.

In this study we explore combining CMNs with a permissioned blockchain that can result in decentralized mesh access networks that make connecting to the internet not only easy and widespread, but trustful and more economical as well.

Our key contributions are summarized as follows:

- First, we deploy the Hyperledger Fabric platform in a production wireless mesh network that is part of Guifi.net. We quantify the performance of the platform in terms of transaction confirmation/completion latency, CPU and memory utlization of HLF components etc. To the best of our knowledge, this is the first Hyperledger Fabric deployment made in a production wireless mesh network.

- Second, driven by the findings in a mesh network, we propose a placement scheme for Hyperledger Fabric components that optimizes the performance of the blockchain protocol.

## 2. BLOCKCHAIN: THE UNDERPINNING TECHNOLOGY

Blockchain is a *append-only* immutable data structure. Its first incarnation was in the Bitcoin cryptocurrency network. Blockchain was used to enable trust for financial transactions among different non-trusting parties in a pure peer-to-peer fashion without the need for going through a third financial party like e.g., a bank. Such trust is provided in terms of immutability of blockchain's data structure. Each *block* in blockchain contains information that is immutable. The immutability aspect is rendered true by including the hash of all the contents of a block into the next block which also chains the blocks together. Tampering of one block disturbs the contents of all the following blocks in the chain. Each block in the chain is appended after a *consensus* is reached among all the peers of the network. The same version of a blockchain is stored in a distributed manner at all the peers of the network. That is why it is sometimes referred to as *distributed ledger* as well.

#### *2.0.1 Open and public blockchain*

Blockchain of Bitcoin [10], Ethereum[4] [15], and in general of various other cyptocurrencies are mostly open and public. This means that anyone can be a part of the blockchain's network and make transactions with other parties. Anonymity is also at the heart of such platforms. A user (or in general an entity) usually uses the hash of its public key as its identifier as opposed to using its real-world credentials. In the aspect of openness the *permissioned blockchains* are in sharp contrast with public blockchains which we discuss next.

#### *2.0.2 Permissioned blockchain*

Permissioned blockchain, a concept particualrly popularized by the Linux Foundation's Hyperledger, are usually considered for business applications. In such applications the identity of users are also important such as the requirement of *know your customers* for many businesses. Hyperledger tries to leverage the best of both worlds by implementing a cryptographic *membership service* on top of blockchain's trusted, immutable, and distributed record keeping. In our study the requirement of both users' identity and trusted record keeping is of paramount importance and that is why we decided to conduct our study using Hyperledger Fabric, which we discuss next.

### 2.1 Hyperledger Fabric (HLF)

Hyperledger Fabric (HLF) is an open source implementation of a permissioned blockchain network that executes distributed applications written in general-purpose programming languages (e.g., Go, Java etc) [2]. HLF's approach is modular, which implies that the platform is capable of supporting different implementations of its different components (such as different consensus protocols) in a *plug-and-play* fashion.

The HLF architecture comprises of the following components:

**Peers:** Peers can further be of two types namely *endorsers* and *committers*. A peer is called a *committer* when it maintains a local copy of the ledger by committing transactions into its blocks. A peer assumes the role of an *endorser* when it is also responsible for simulating the transactions by executing specific chaincodes and endorsing the result (see the next subsection 2.2). A peer can be an endorser for certain types of transactions and just a committer for others.

**Ordering service:** The role of this component is to order the transactions chronologically by time stamping them to avoid the *double spend problem* [10]. The ordering service create new blocks of transactions and broadcast them to the peers which then append these blocks to their local copy of the blockchain (or ledger). The ordering service can be implemented as a centralized or decentralized service [14]. It is at the ordering service where the actual consensus (like proof-of-work in Bitcoin [10]) takes place.

[3] `https://www.hyperledger.org/`

[4] `https://ethereum.org/`

**Chaincode:** A chaincode or a *smart contract* is a program code that implements the application logic. It is run in a distributed manner by the peers. It is installed and instantiated on the network of Hyperledger Fabric peer nodes, enabling interaction with the networks shared ledger (i.e., the state of a database modeled as a versioned key/value store).

**Channel[5]:** A channel provides a higher layer of confidentiality abstraction. A channel can be considered as a subnet on top of a larger blockchain network. Each channel has its own set of chaincodes, member entities (peers and orderers), and a distinct version of a distributed ledger.

## 2.2 HLF Protocol

Figure 1 depicts the sequence of transaction execution steps in HLF's environment. The description of these execution steps are as follows:

**1. Tx proposal:** In this step clients access the HLF blockchain to submit a proposal for a Tx to be included in one of the blocks of the HLF blockchain. Clients propose a transaction through an application that uses an SDK's (Java, Python etc) API. This is shown as the first step in Figure 1.

**2. Endorsement and Tx simulation:** The transaction proposal from the above step is then broadcasted to the endorsing peer nodes in the HLF blockchain network. Each endorsing peer verifies the Tx proposal in terms of its correctness (i.e., its structure, the signatures that it contains, and the membership and permission status of the client that submits the transaction) uniqueness (i.e., this proposal has not be submitted in the past).

After the above checks comes the *transaction simulation step*. Endorsing peers invoke a relevant chaincode (as specified in the Tx proposal by the submitting client). The execution (as per specific arguments specified in Tx proposal) of this chaincode produces an output against the current state of the database (ledger). Without updating the ledger's state, the output of the Tx simulation is sent back in the form of *proposal response* back to the client through the SDK. In Figure 1 this is shown by the second step.

**3. Inspection of proposal response:** After the above step the client-side application collects the responses from the endorsement step. Afterwards all the responses are cross checked (in terms of the signatures of the endorsing peers and the content of the responses) to determine if there are any disparities among the content of the responses. If the content of all the responses are the same and according to the pre-defined *endorsement policy* (i.e., number of peers whose endorsements—in terms of their signatures—are necessary) then the client submits this Tx to the Ordering Service (more on it in the next step) that will in turn ultimately update the ledger's state as per the Tx simulation outcome in the last step.

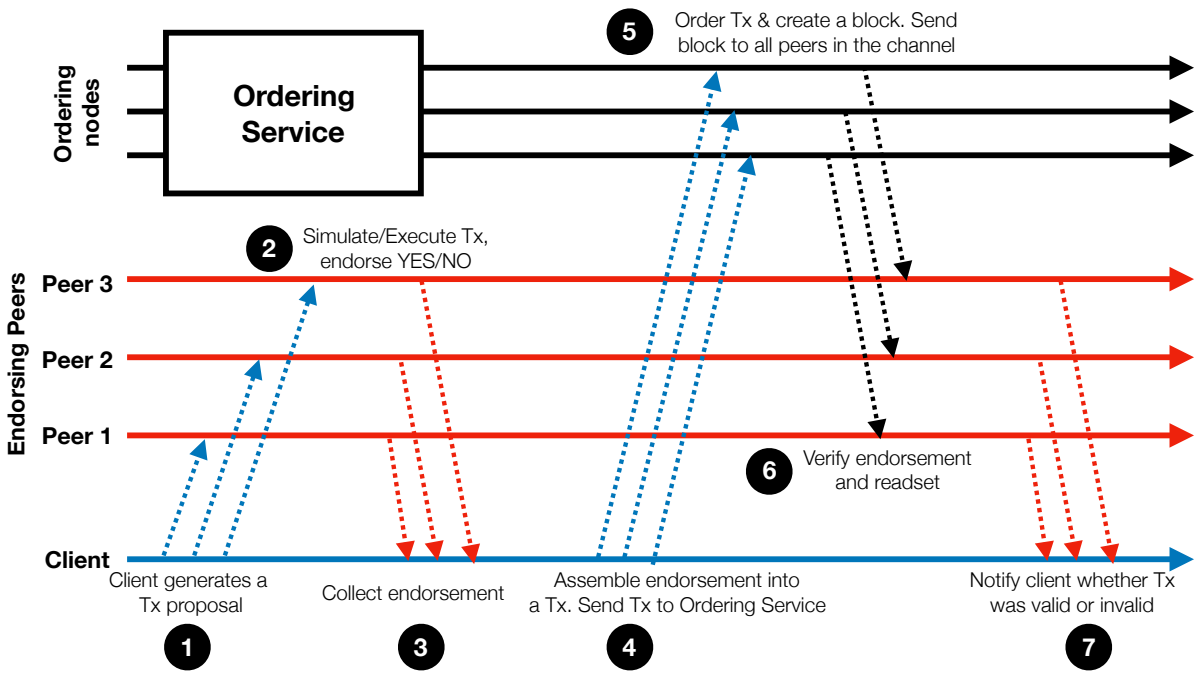

Figure 1: Hyperledger Fabric Protocol

It can also happen that in the Tx proposal, made in the last step, only the current state of the ledger was queried. In this case there will be no need to update a ledger's state and hence there is no submission to the Ordering Service by the client. In Figure 1 this is shown by step three.

**4. Tx submission to the Ordering Service:** The Ordering Service collects various Txs after the last step via various channels. It then *orders* them according to their receiving times at the Ordering Service. This ordered set of Txs is then included in a block specific to a channel which will later be appended in the channel's ledger. Step four and five in Figure 1 shows this process.

**5. Tx validation and commit:** In this stage all the peers belonging to a particular channel receive a block containing Txs specific to this channel. Each peer then checks all the Txs in terms of their validity. Valid Txs are those that satisfy an endorsement policy. If the Txs pass the validity test then they are tagged as valid otherwise invalid in a block and then this block is appended to the ledger maintained by the peers of this channel. This is coverd by step six in Figure 1.

**6. Ledger update notification:** Finally, after the ledger update in the last step the client of the submitting Tx is notified about the validity or invalidity of the Tx that was included in the latest block of the channel's distributed ledger. This is step seven in Figure 1.

# 3. CASE STUDY: QMPSU MESH

Quick Mesh Project (qMp) [6] provides a firmware based on OpenWrt Linux with the aim to ease the deployment of mesh networks by the users who are willing to interconnect in an area, and pool their Internet uplinks [6]. qMp was initiated in 2011 by a few Guifi.net activists.

[6] `http://qmp.cat/Overview`

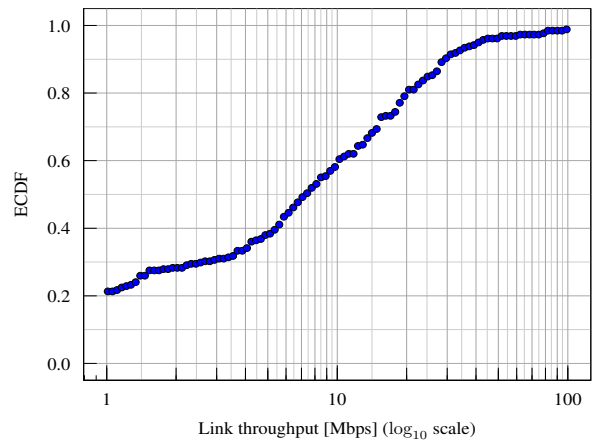

Figure 2: Bandwidth ECDF

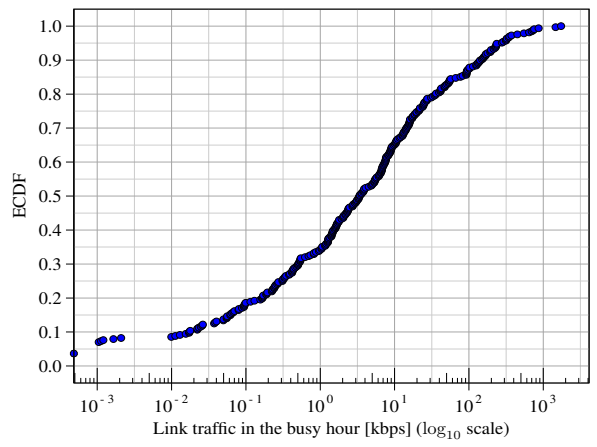

Figure 3: Traffic ECDF

The qMp firmware has enabled to deploy several mesh networks with actual end-users (e.g., more than 250 active locations, typically households) in several parts surrounding the city of Barcelona[7]. At the time of this writing, there are 10 different mesh networks, and the largest (Sants-UPC or QMPSU) has 85 operational nodes. In that network, there are two gateways that connect the QMPSU network to the rest of Guifi.net and the Internet. Users join the mesh by setting up *outdoor routers* (i.e., antennas) that automatically establish router-to-router links. The outdoor routers are connected through Ethernet to an indoor AP (access point) as a premises network where the edge devices and services are running: home-servers such as Raspberry Pi's or Cloudy devices [4].

**Network Performance:** We monitored the QMPSU mesh network for a period of one month. We took hourly captures from the network for the entire month of March 2018. Figure 2 and 3 depict the bandwidth and traffic distribution of all the links in the network. Figure 2 shows that the link throughput can be fitted with a mean of 13.6 Mbps. At the same time Figure 2 reveals that 60% of the nodes have 10 Mbps or less throughput. Figure 3 demonstrates that the maximum per-link traffic in the busiest hour is 1736 kbps. We observed that the resources are not uniformly distributed in the network. There is a highly skewed bandwidth and traffic distribution.

**Node Deployment:** Based on the network measurement analysis we have strategically deployed 10 Raspberry Pi (RPi3) devices on the outdoor routers to cover the area of the QMPSU network as presented in Figure 4. We use our previous work [13] on service placement to determine nodes in the network. In this set, we cover nodes with different properties: with higher bandwidth [13], nodes that are highly connected (i.e., with high degree centrality) [7], nodes acting as bridges (with high betweenness centrality) and nodes not well connected. After the nodes were chosen, we deployed 10 RPi boards in the community users home.

# 4. EVALUATION

We setup a blockchain testbed network comprising RPi3 boards, each running a component of Hyperledger Fabric (HLF) in the QMPSU network. In this testbed, different RPi3 nodes run different components of HLF (see Section 2.1 for details on HLF components). In parallel, we also deploy a similar setup in the lab environment (for performance comparison purposes) and evaluate the performance in both environments.

[7]`http://dsg.ac.upc.edu/qmpsu`

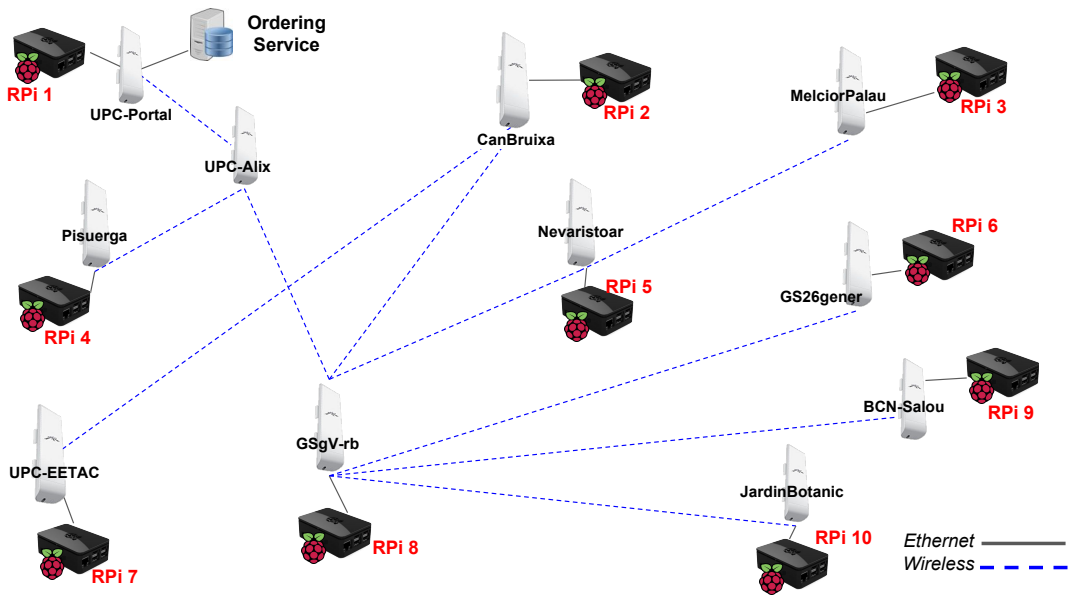

Figure 4: Topology of the deployed nodes

## 4.1 Experimental setup

In our experiments, we deploy a HLF blockchain network consisting of a single organizational entity. All the transactions happen among the members of this single organization. The HLF components, namely peer (we deploy multiple instances of this component), orderer, and client are deployed in different RPi3 boards connected to each other in the same local network. The RPi3 boards have 1.2GHz 4 core ARM cortex A53 processor, a RAM memory of 1GB and run *raspbian-stretch* OS. Both, in the lab and in the QMPSU network, we performed experiments by placing different HLF components at different physical (RPi3) nodes and by varying the number of peers from 1 to 4. We evaluate the setup in the lab and in QMPSU network by comparing transaction latency of HLF when 100 transactions are fired serially and in parallel. We also evaluate transaction latencies in HLF for a 2 peer setup when the block size is varied from 10 to 100 transactions per block. Our experiments comprise of 3 runs (taken in different time slots) and the presented results are averaged over all the runs.

## 4.2 Results

### *4.2.1 Transaction Latency*

Table 1 lists the transaction completion time (referred to as *TTC: Time-to-Commit*) for 100 transactions, initiated in parallel, between the two peer nodes in the lab environment and in the QMPSU network respectively with block sizes ranging from 10 to 100 transactions per block. It can be observed that, as the block size increases, the transaction completion time increases both in the lab setup as well as in the QMPSU network.

Transaction latency is defined as total time taken to endorse and to commit a transaction to the ledger. Figure 5 shows the comparison of transaction latency observed for two different placements of HLF ordering service. We measured the transaction latency when the HLF ordering service is placed randomly in the network (Random) and when it is placed at the node chosen with a heuristic that considers the node with higher bandwidth and degree centrality (BASP) [13]. The results of Figure 5 are obtained when a client initiates 100 transactions sequentially. This Figure reveals that the gain brought by BASP , for the case when we have one endorser in the network, is 30.8%. For the case when we have four endorsers in the network, the gain of BASP over Random is 24%. Further, Figure 5 demonstrates that in the QMPSU network it takes 1.2 seconds for a single transaction to be appended to the distributed ledger.

| Block Size | TTC(Lab) | TTC(QMPSU) | # of Tx |
|---|---|---|---|
| 10 | 33.4 s | 64.2 s | 100 |
| 20 | 35.0 s | 69.7 s | 100 |
| 50 | 39.2 s | 75.3 s | 100 |
| 100 | 45.3 s | 84.8 s | 100 |

Table 1: Transaction delivery time (parallel transactions)

### 4.2.2 *Resource Consumption*

Figure 6 shows CPU utilization by various components of the HLF network namely: an orderer, a client and two peers (an endorser and a committer). CPU utilization of all nodes is monitored for a time period of 60 seconds during which 100 transactions are initiated in parallel (by the client) and all the transactions are completed. 100 parallel transactions took around 40 seconds to complete. We chose to monitor the nodes for a time period of 60 seconds to show idle phase usage and busy phase usage of each node. In the graph, transactions are fired at 11th second and all the transactions get completed at 50th second. It can be observed that the endorser is the node with the highest CPU utilization whereas the orderer utilizes the least of CPU.

The Figure 6 shows that, for 100 transactions initiated at the same time, the endorser's maximum CPU utilization reaches 96%. The maximum CPU utilization is 81% for the committer while it is 71% for the orderer. The reason that the endorser has the highest CPU consumption, among other HLF components, is because of the chaincode execution at the endorsing peer, which does not happen in the committer and the orderer.

The chaincode container executes the chaincode for each incoming transaction which does not happen at the committer node. When multiple transactions happen in parallel, concurrent execution of the chaincode happens for all transactions, increasing the load on the endorsing peer. With 100 parallel transactions, we observe that the CPU load reaches to 96% at the endorser. However, the load on each endorser can be reduced by deploying multiple endorsers in the network. The load on different endorsers can be balanced by designing a suitable endorsement policy and devising a strategy at the client to request endorsements from different set of endorsers each time a transaction is initiated.

Similarly, memory usage is the highest by the endorser and the least by the orderer. Memory usage of committing peer falls in between of endorsing peer and the orderer. In orderer and committing peer, memory usage remains almost the same between the idle phase and during transaction execution. Memory usage in orderer mostly falls in the range of 57%-58% while the memory usage in the committer is in the range of 57%-60%. In endorsing peer the memory usage increases during transaction execution as the execution of a chaincode also takes place at the same time. The memory usage by the endorser is about 60% in idle phase and reaches to a maximum of 65% during the chaincode execution.

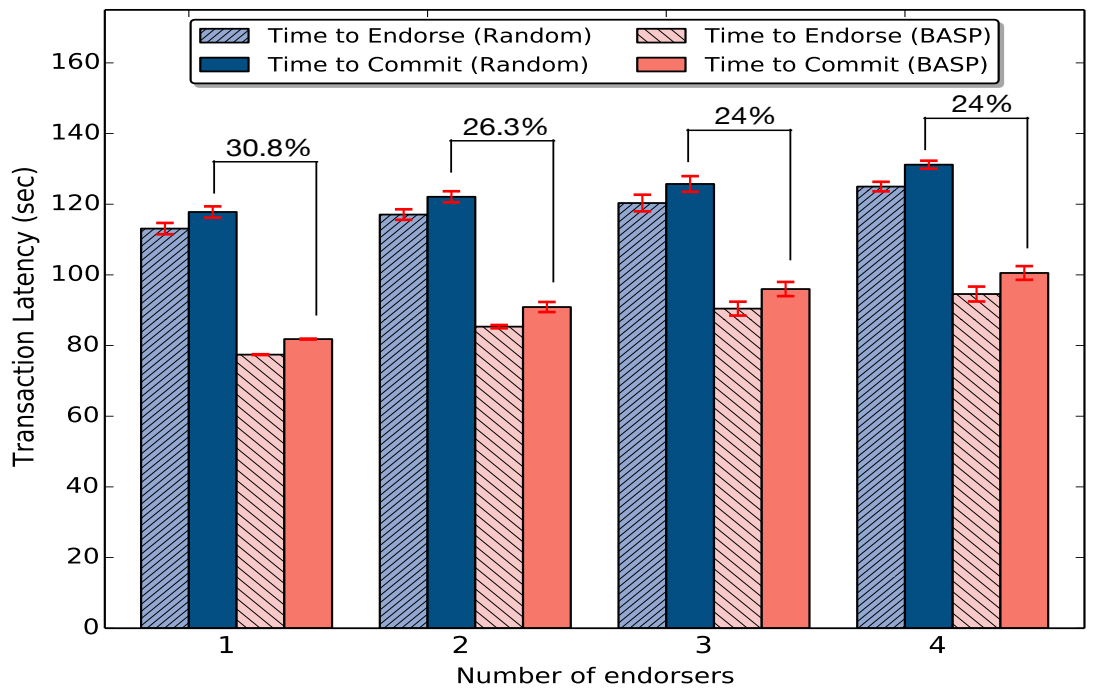

Figure 5: Transaction latency (QMPSU)

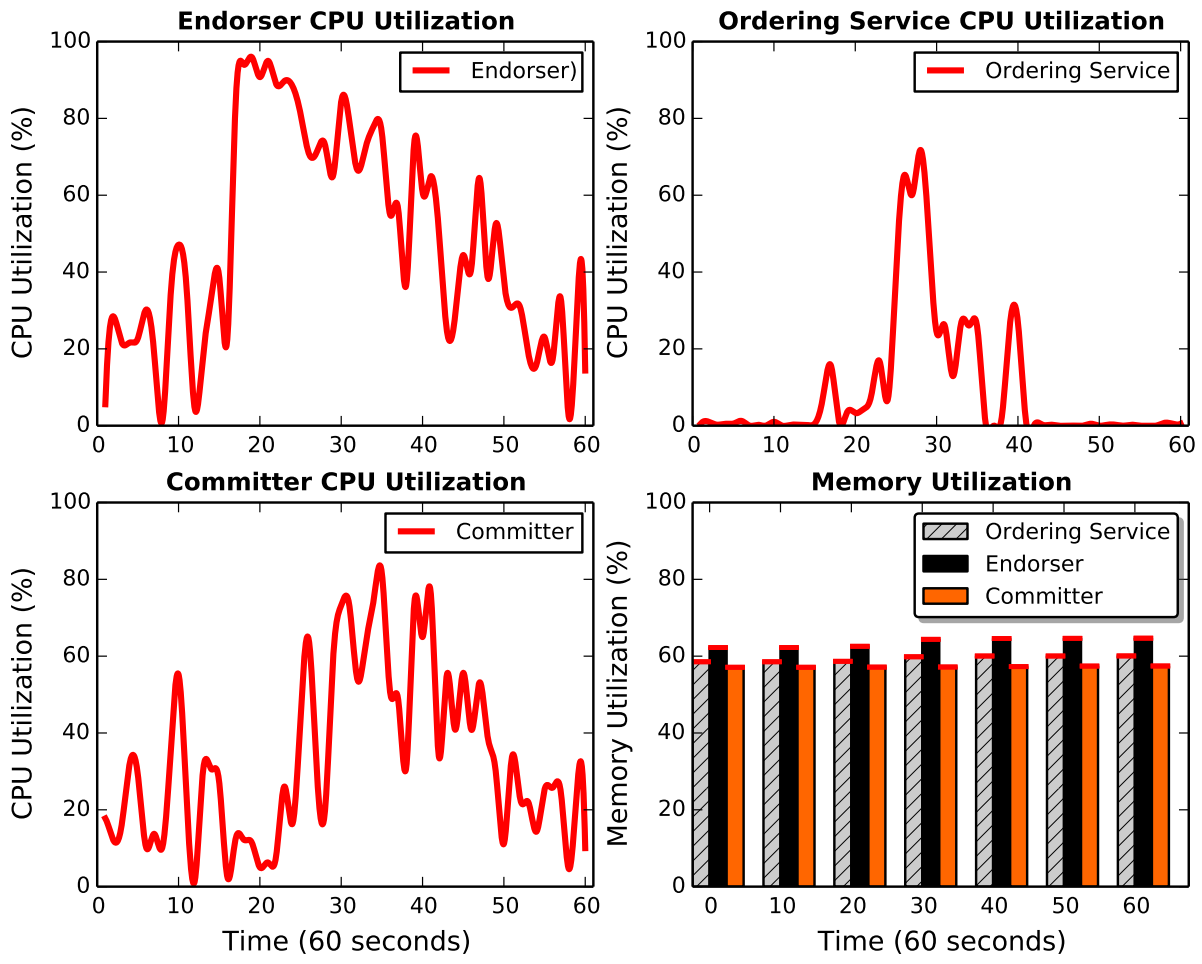

Figure 6: CPU and memory utilization

## 4.3 Discussion

As we observed in our experiments that, in terms of resource consumption, the endorser nodes can prove to be a bottleneck. We believe that this bottleneck is because of the execution of an additional chaincode container at each endorsing node. In our current study we only considered one endorser node to study the resource utilization with a simple endorsement policy encoded in the corresponding chaincode. It might get more complicated when we consider more than one endorsers and with a sophisticated endorsement policy (however, as discussed in Section 4.2.2, if done right it can actually improve performance). In addition to this, the actual distribution of endorsing peers in a production network, such as QMPSU, might also affect the network performance (both in terms of CPU utilization and transaction latency). So care must be taken, specially in the resource constrained nature of CMNs, in designing an endorsing policy that is also cognizant of the underlying network infrastructure (i.e, its topology; is it mostly wired or wireless? etc).

## 5. RELATED WORK

The study [12] compares the public blockchain with permissioned blockchain and discusses the trade-offs among decentralization, scalability and security in the two approaches. Sousa et al. [14] present the design, implementation and evaluation of a BFT ordering service for Hyperledger Fabric based on the the BFT-SMART state machine replication/consensus library. Their results show that Hyperledger Fabric with their ordering service can achieve up to ten thousand transactions per second and write a transaction irrevocably in the blockchain in half a second, even with peers distributed over different continents. The Blockbench [8] is framework for analyzing private blockchains. It serves as a fair means of comparison for different platforms and enables deeper understanding of different system design choices.They use Blockbench to conduct comprehensive evaluation of three major private blockchains: Ethereum, Parity and Hyperledger Fabric. Their result demonstrate that these systems are still far from replacing the current database systems in traditional data processing workloads. Most of the above mentioned works are not done in CMNs context and are not applicable to our scenario.

## 6. CONCLUSION

The missing ingredient for widespread adoption of CMNs has always been the issue of economic sustainability. In this paper, we take on the issue of addressing trustworthy economic sustainability by proposing the need for an economic substrate built using blockchain that can keep a record of the transactions related to the contributions (of nodes, links, Internet gateways, maintenance) and consumption of communication network's resources in a decentralized and trusted manner. The evaluation of the Hyperledger Fabric blockchain deployment in the laboratory and in a real production mesh network gives us an understanding of the performance, overhead, influence of the underlying network, and limitations of this framework. The results show critical aspects that can be optimized in a Hyperledger Fabric deployment, in the perspective of CMNs, where several components can prove to be bottlenecks and therefore put a limiting effect on the rate of economic transactions in a mesh network. Future work will expand the evaluation to a wider range of hardware and network configurations and considering real and synthetic transaction traces with a more realistic design of an endorsement policy (chaincode).